\documentclass[fleqn,12pt]{wlscirep}
\newcommand{\FD}[1]{\textcolor{black}{#1}}
\newcommand{\AL}[1]{\textcolor{black}{#1}}

\usepackage[utf8]{inputenc}
\usepackage[T1]{fontenc}

\usepackage{subfig}

\title{Insights into nanoparticle shape transformation by energetic ions using atomistic simulations}

\author[1]{Aleksi A. Leino}
\author[1]{Ville E. Jantunen}
\author[1]{Flyura Djurabekova}
\affil[1]{University of Helsinki, Department of Physics, Helsinki, P.O. Box 43, FI-00014, Finland}

\affil[*]{aleksi.leino@helsinki.fi}

\begin{abstract}
The shape of metal nanoparticles embedded in dielectric matrices influences the optical properties of the composite material. Swift heavy ion irradiation can induce a controllable shape transformation in gold and other metals embedded in amorphous silicon dioxide, where the particles elongate along the direction of the ion beam. The details of this transformation are not fully understood, but it is presumably related to nanometer-scale phase transitions induced by individual ion impacts. The phenomenon has been reproduced using atomistic simulations, \FD{although the time scale limitations and the lack of accurate interatomic models within the metal-silica interface lead unavoidably to} 
severe simplifications. We improve the realism in the simulations with an accurate model for surface adhesion \AL{between gold and silica} and by simulating the processes in the matrix between impacts. The simulations with correct adhesion show that the nanoparticles can grow in aspect ratio in the molten state even after silicon dioxide solidifies. Moreover, we demonstrate the active role of the matrix: without explicitly modeling processes in the matrix between impacts, the elongation \FD{is limited and does not reach significant aspect ratios seen in experiments.} 
These results \AL{significantly improve} the theoretical understanding \FD{of processes developing in embedded nanoparticles under swift heavy ion irradiation. The knowledge brings forward the ion beam technology as} 
a precise tool for \FD{shaping 
of embedded nanostructures for various optical applications. }

\end{abstract}
\begin{document}
\flushbottom
\maketitle

\thispagestyle{empty}

\section*{Introduction}
Ion irradiation is a versatile tool for analyzing and modifying materials.  When ions are accelerated above specific kinetic energies exceeding about 1 MeV/u, the interaction with the electrons in the material becomes the primary cause for the deceleration of the ion. Ions in this energy regime are called swift heavy ions (SHIs).  The characteristics differ from that of the keV -regime. Most of the effects from individual ions are confined within a cylindrical region of several microns in length but only some nanometers in diameter.

SHI irradiation causes a peculiar shape transformation induced to nanoparticles (NPs) embedded in amorphous $\text{SiO}{}_2$ matrices. Numerous experiments\cite{DOrleans03, Roorda04, SilvaPereyra10, Rizza12, Awazu09, Kluth09,Ridgway11, Leino14, Coulon2016, MotaSantiago17,  Amekura19, Li2020} have shown that spherical NPs can become elongated in the ion beam direction and evolve into prolate shapes or rods. While intriguing from an application point of view, the shaping process also offers unique insights into the fundamentals of ion-solid interactions. 
Consequently, the phenomenon has been extensively researched\cite{Leino14b} since the first reports\cite{DOrleans03}. From an application point of view, the process produces large arrays of equally aligned nanoparticles embedded within a solid. It has technological relevance since, for example, the shape of the NPs -- controlled by the irradiation parameters -- affects the collective response of the composite material to external electromagnetic radiation. This control is essential in plasmonics research, whose potential applications range from sensing and imaging devices\cite{Murphy08} to building subwavelength photonic circuits\cite{Barnes03}. However, no complete theoretical description of the shaping process has been developed so far, and even the primary mechanism by which the transformation occurs is still subject to debate. 

Building a theoretical description has been particularly challenging as many details related to the swift heavy ion collision kinetics and the relaxation pathways of the initial, highly excited electronic subsystem are still not well understood. Quantitative analysis usually invokes the two-temperature model (inelastic thermal spike model), Coulomb explosion model, and the exciton self-trapping model to explain the experimental observations\cite{Itoh2009}. Due to the empirical parameters present in the models, it has not been possible to distinguish between them. While most first-principles approaches\cite{Correa12, Ojanpera14} are still limited to studying femtosecond dynamics, Monte Carlo\cite{Gervais94,Murat08} simulations describing the initial kinetics of the electrons and holes can reach longer time and spatial scales. Recently, these have been combined with molecular dynamics (MD) simulations\cite{Rymzhanov20}. Such approaches are promising, but the number of cases studied extensively is still low. Moreover, describing relevant phases accurately with the interatomic potentials and comparing the modeling results to experiments alone poses a formidable challenge. Regardless, SHI effects can often be successfully explained by rapid heating of the lattice in cylindrical symmetry irrespective of the underlying theoretical model invoked to explain the heating\cite{Trinkaus1995, Klaumunzer06}. Consequently, simulating rapid, cylindrical heating in classical MD simulations provides a powerful tool to understand the later stages of a swift heavy ion impact\cite{Kluth2008, Vazquez2021}, even without a first-principles description for the initial radial distribution of kinetic energy.

{Accordingly, the elongation phenomenon has been mostly explained in terms of heating effects. Two prominent models have emerged along this vein. The first considers the ion-hammering effect, which, due to the combined effect of many transient, molten ion tracks,  causes the matrix to strain perpendicularly to the beam\cite{Trinkaus1995,Klaumunzer06}. This effect occurs above a threshold electronic stopping power and fluence in amorphisable materials and generates perpendicular stresses within the NPs \cite{Klaumunzer06}. These stresses have been speculated to shape the nanoparticle\cite{Roorda04, Kerboua13, Coulon2016} when combined with the irradiation-induced softening or melting of the NP. The second model considers the pressure, volume expansion, and confinement of molten gold on impact\cite{DOrleans03, Klaumunzer06, Leino14, PenaRodriguez17}. A small elongation increment occurs after each impact. In a scenario supported by MD simulations\cite{Vu19, Leino14}, the molten and expanding material from the NPs pushes to the underdense, molten track on top and beneath the nanoparticle on impact. Silica remains solid elsewhere when the ion trajectory intersects the NP from the center region. On the other hand, it has been calculated that impacts to the sides of a sufficiently large NP do not heat it above melting point\cite{Dufour12, Coulon2016}, explaining why the elongation process occurs in the beam direction. Additionally, dissociation of the nanoparticle on impact\cite{DOrleans03, Roorda04, PenaRodriguez17} and spin-spin interactions\cite{Sarker16} have been considered to contribute to the shape change.
}
The ion hammering scenario gains support by the observation that when the metal NPs are embedded in small colloidal silica particles, they do not elongate\cite{Roorda04}. NPs in larger colloids than 26 nm in thickness, however,  elongation. However, as contrary evidence, a small shape change was observed below the threshold fluence for ion hammering, indicating deformation by individual impacts\cite{amekura2017}. Moreover, theoretical continuum mechanical considerations do not support the idea of deformation by hammering\cite{Klaumunzer06}. Recent works include the observation of elongation using C${}_{60}$ ions\cite{Amekura19}, studies on embedded NPs at the interface of silica and silicon nitride\cite{MotaSantiago17}, and the observation of elongation in NPs embedded in crystalline indium tin oxide \cite{Li2020}. See for example,  Refs.\cite{Rizza15,Leino14b,Chen2020} for more complete reviews on the elongation phenomenon.

Atomistic simulations performed by several groups have shown that NPs change shape on impacts\cite{Leino14, PenaRodriguez17, Vu19, Amekura20}. On the other hand, the mere reproduction of the effect in the simulation does not unambiguously identify the underlying mechanism, and the conclusions drawn from MD simulations appear contradictory. The authors in Ref.\cite{PenaRodriguez17} attribute the elongation to the formation of a shock wave, extremely high temperatures of the NP (9000K), and associated interfacial pressures. On the other hand, other authors\cite{Leino14, Vu19, Amekura20} contribute the driving force to be the volume expansion due to melting and heating that occurs in considerably lower temperatures (2000K). Using first-principles MD computations, authors in Ref.\cite{Sarker16} suggested that spin-effects play a central role in the elongation. Additionally, the effect of interparticle diffusion has been investigated using the lattice kinetic Monte Carlo method\cite{Khan18}.

In this article, we study the role of the silica matrix using classical MD simulations. We introduce a realistic description of Au-SiO$_2$ interactions using density functional theory (DFT) and show that the aspect ratio of the NP can change even after the ion track has solidified due to the adhesive forces between gold and silica. Furthermore, the evolution of the NPs between subsequent impacts was either neglected or heavily simplified in previous works. Here we demonstrate that secondary processes, i. e. indirect impacts, and relaxation of the NP, play a significant role in the shape transformation. 
We present the results of long-scale MD simulations that reveal an amorphous-to-crystalline phase transition and the build-up of negative stress in the NP within the first six ns after the impact. We show that this stress is relaxed by indirect impacts, which allows the elongation process to continue. The results suggest that the previous explanation models were oversimplified: both the NP and the matrix have an active role in the shape transformation.

\section*{Results and discussion}

\subsection*{Interatomic potential for the Au-silica system}
Previous studies on the amorphous silica-Au system using first-principles simulation methods have revealed a high diffusion barrier for gold through silica rings\cite{Ulrich09}, weak binding\cite{DelVitto05, Huhn15, Ulrich09} to regular sites, surface dewetting\cite{Huhn15}, and that the interface does not intermix\cite{Huhn15}. In the first simulations to reproduce the elongation phenomenon\cite{Leino12, Leino14} with Au NPs, a universal, purely repulsive pair potential between gold and silica was used to give similar characteristics to the interface as a first approximation.

\begin{figure}[!h]
\centering
\includegraphics[width=0.6\linewidth]{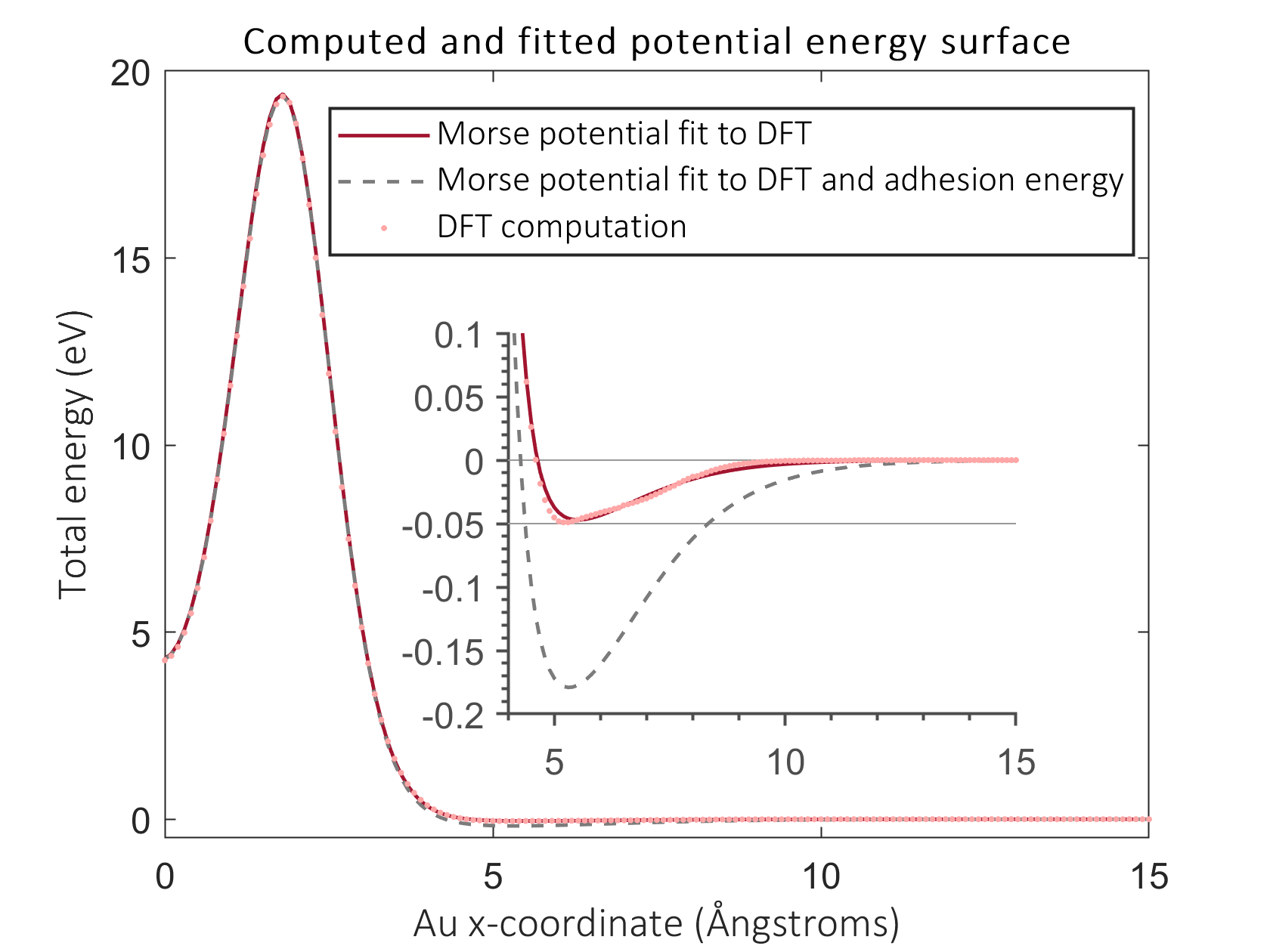}
\hspace{-5mm}
\includegraphics[width=0.4\linewidth]{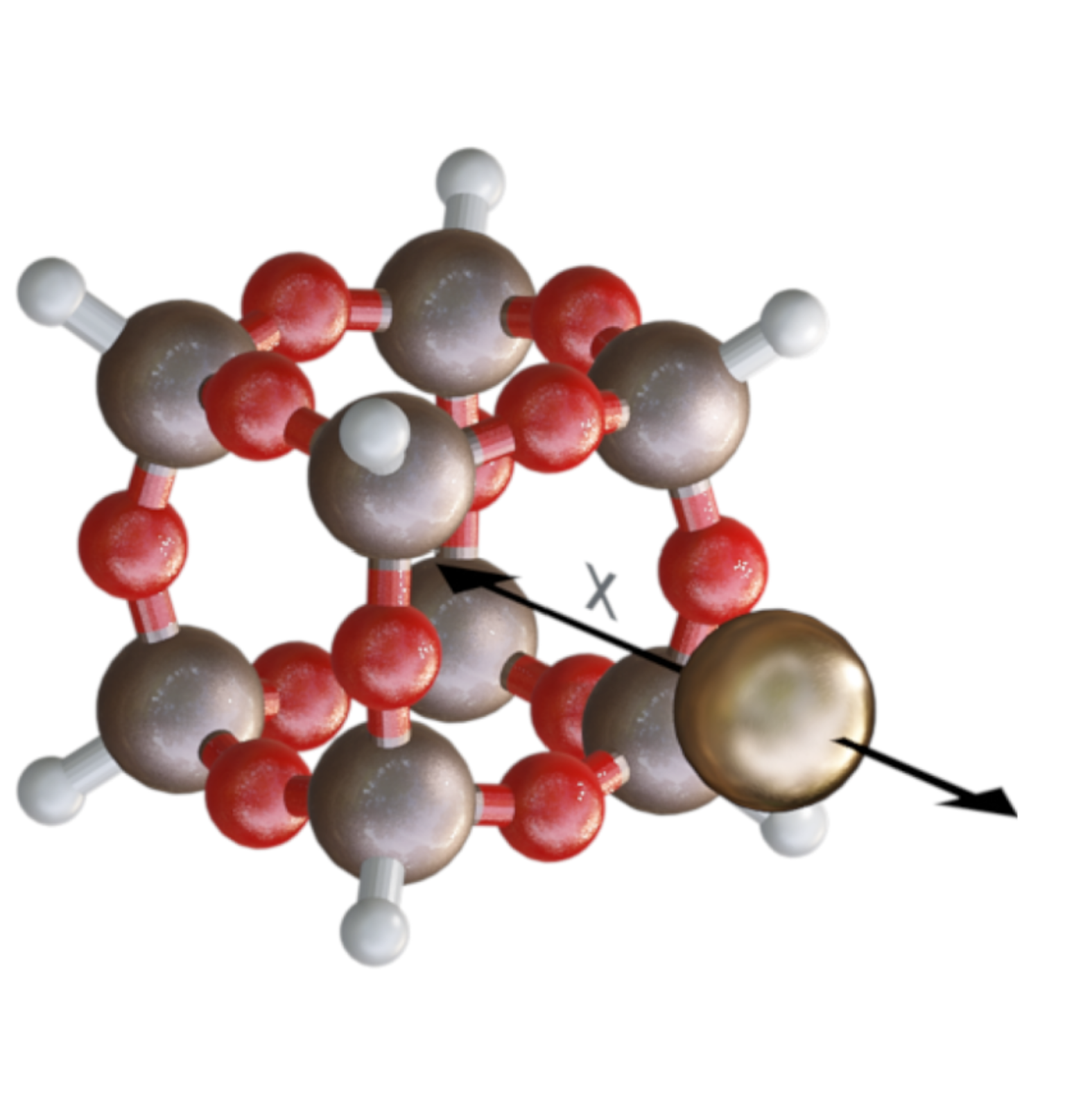}
\caption{Potential energy surface scan for the Au-
H${}_{\text{8}}$Si${}_8$O${}_{12}$ system.
The graph on the left shows the data points obtained from the scan, and the Morse potential fits. The solid line indicates the original potential and the dashed shows the potential after fitting to adhesion energy as explained in the methods. The figure on the right shows the geometry of the relaxed molecule and illustrates the x-coordinate. x=0 points to the middle of the molecule.
 }
\label{fig:pes_scan}
\end{figure}
As pointed out by Klaümunzer\cite{Klaumunzer06}, the interfacial energies (energy cost to form the surface) in metal-ceramics are in the order of 1 J/m${}^2$, which can cause interfacial pressures in metal NPs comparable to those from the ion hammering effect. The adhesive energy (the cost to separate the surfaces) of the Au-silica surface is several hundreds of mJ/m${}^2$ in magnitude\cite{Gadkari05, Huhn15} but not previously taken into account. Hence, we introduce a Morse-type potential that describes the interface adhesion and the diffusion barrier through silica rings by fitting it to the energies predicted by a DFT computation and the experimental value of adhesive energy. Following Ref.\cite{DelVitto05}, we base our pair potential on a octasilasesquioxane (H${}_{\text{8}}$Si${}_8$O${}_{12}$) -molecule, shown in Fig. \ref{fig:pes_scan}, to reduce the computational cost.  

\begin{table}[!ht]
\centering
\caption{\label{tab:potential} 
Fitted Morse potential $V(r) = D_e (e^{-2 a (r - r_e) } - 2e^{a (r - r_e})$ parameters for Au-Si, and Au-O interactions. Shown is the parameters for the original potential and the one with corrected adhesive energy (see Fig. \ref{fig:pes_scan}).
}
\begin{tabular}{|l|l|l|}

\hline
Parameter & Original potential  & Corrected potential \\
\hline
$D_{e,\text{Si-Au}}$ & 0.3163 eV &  0.4260 eV\\
\hline
$a_{\text{Si-Au}}$ & 1.8536 1/Å & 1.8139 1/Å \\
\hline
$r_{e, \text{Si-Au}}$ & 2.6327 Å &  2.5816 Å \\
\hline
$D_{e,\text{O-Au}}$ & 0.0014 eV & 0.0091 eV \\
\hline
$a_{\text{O-Au}}$ & 0.9082 1/Å & 0.8175 1/Å\\
\hline
$r_{e, \text{O-Au}}$ & 6.3961 Å & 5.8094 Å \\
\hline
Adhesive energy & 36 mJ/m${}^2$ & 430 mJ/m${}^2$ \\
\hline

\end{tabular}
\end{table}

We use hydrogen atoms to terminate the bonds of Si atoms so that the structure represents a defect-free region of silica. We obtain the Morse potential by fitting it to the potential energy curve obtained by dragging a gold atom through a silica ring, as indicated in Figure \ref{fig:pes_scan}. The curve shows that gold atom binds to the surface of the molecule with a binding energy of about 50 meV. However, it is well-known that regular DFT underestimates the strength of dispersion interactions. Authors in Ref. \cite{Huhn15} used the Grimme dispersion correction to DFT to compute the adhesive energy of the silica-Au interface, which showed that for hydroxylated surfaces, the adhesion consists almost entirely of the dispersion correction. In line with this result, we calculate that the resulting adhesive energy without any corrections is approximately 40 mJ/m${}^2$, whereas the experimental value for hydroxylated surfaces is 360 mJ/m${}^2$ at room temperature\cite{Kwon2003}. To correct the adhesion predicted by the potential, we deepen the attractive well to yield adhesion energy of 430 mJ/m${}^2$, as shown in Fig. \ref{fig:pes_scan}. The potential obtained this way also suits better for the intended use at Au-SiO${}_2$ surface rather than with isolated Au atoms.

Parameters for the original and adhesive energy-corrected potential are given in Table \ref{tab:potential}. We note that the experimental value for the adhesive energy (360 mJ/m${}^2$) is for measured hydroxylated glass surfaces. However, silica may contain unpassivated defect sites depending on the experimental conditions. They can form over 2 eV bonds with gold atoms\cite{DelVitto05} and increase the adhesive energy\cite{Huhn15}. Therefore, even the corrected adhesive energy (430 mJ/m${}^2$) should be considered an underestimation. 

\begin{figure}[!ht]

\subfloat[]
{
  \includegraphics[width=0.65\linewidth]{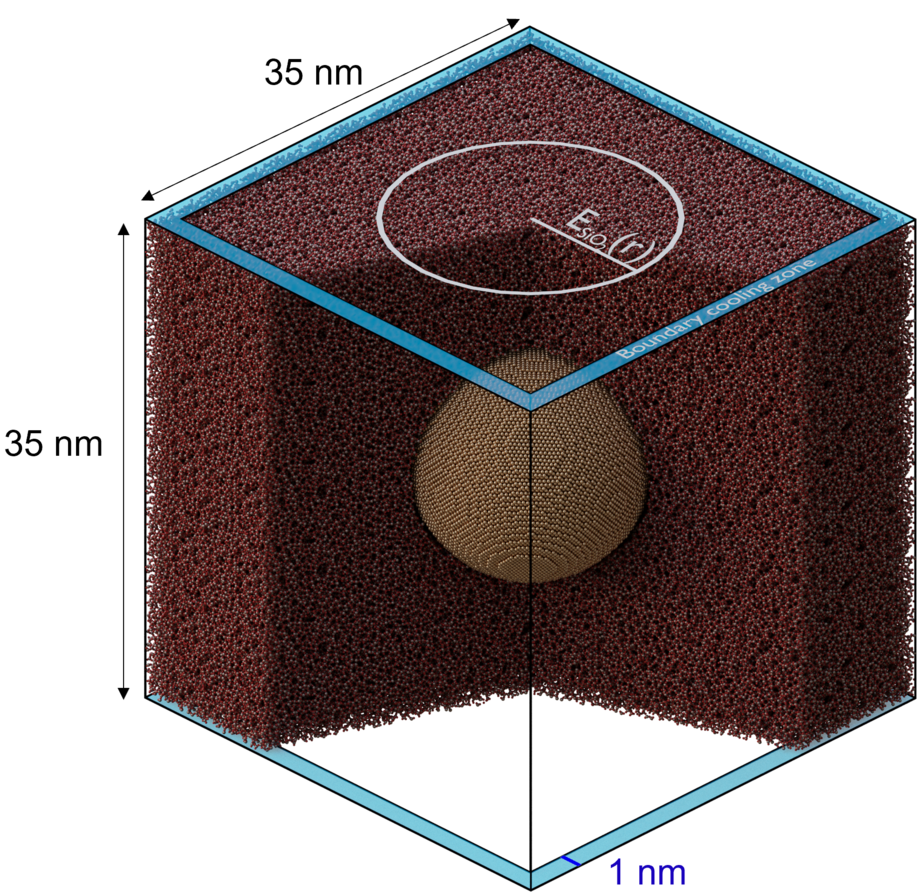}
   \label{fig:fig2}
}
\subfloat[]
{
    \includegraphics[width=0.35\textwidth]{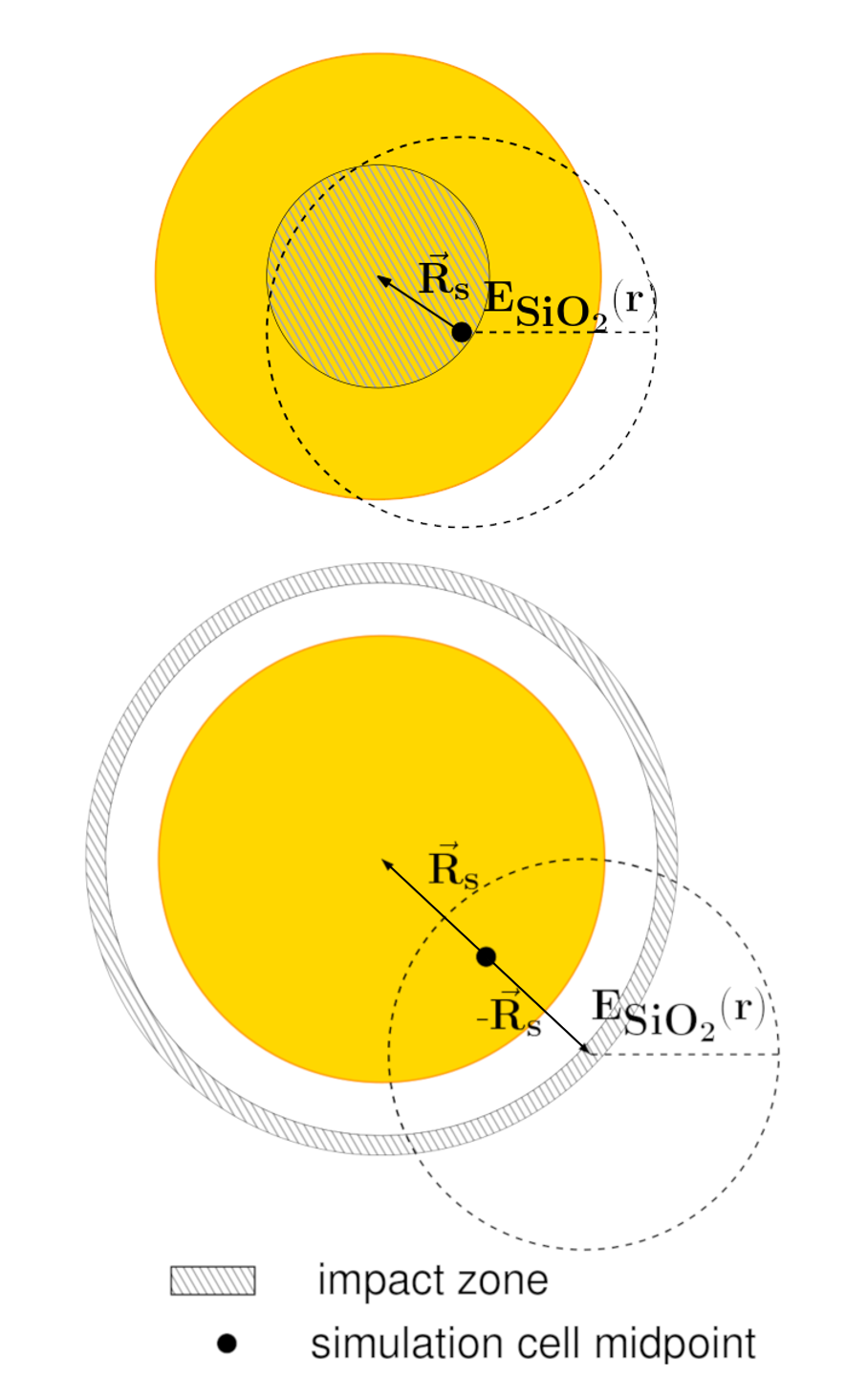}\\
    \label{fig:kirkas}
}
\caption{Schematics of the simulation cell. a) Indicated in the figure 
are the radial symmetry of the energy deposition profile in SiO${}_2$,
cell dimensions, and the width of the boundary cooling volume at the edges of the simulation cell. The radius of the crystalline nanoparticle is 8 nm. b) The upper figure shows the region where ions hit the NP. It also shows an example translation, $\vec R_s$, over the periodic boundary to obtain an impact location within the zone. In this case, $|\vec R_s|_{max} $ = 4 nm. The lower figure shows the translation when the ions hit outside the NP. The radial energy deposition profile is also shifted as indicated. 
The distance from the ion trajectory to the center of the NP is 12 nm, i.e. $|\vec R_s| = $ 6 nm.}

\label{fig:schematics}
\end{figure}

\subsection*{Role of surface adhesion}

To elucidate the role of surface adhesion, we perform two sets of irradiation simulations on spherical NPs with 16 nm diameter using energy depositions estimated for 164 MeV Au ions. The first is based on the potential that \FD{is obtained directly from the DFT calculations without any correction (the adhesion between gold and silicon dioxide is underestimated, $E_{adh}=$36 mJ/m${}^2$),} while the second includes the adhesion correction ($E_{adh}=$430 mJ/m${}^2$). The simulation cell is depicted in Fig. \ref{fig:fig2}. The path of the ion always intersects the NP, and the impact position is randomly shifted at maximum by half the NP radius between subsequent impacts, as indicated in Fig. \ref{fig:kirkas}. The total duration of simulation of a single impact is \AL{4}00 ps. Temperature control \FD{at 300 K everywhere in the cell at the end of each impact 
allows avoiding} the accumulation of energy in the cell.

\begin{figure}[ht!]
\centering
\subfloat[]
{
   \includegraphics[width=0.5\textwidth]{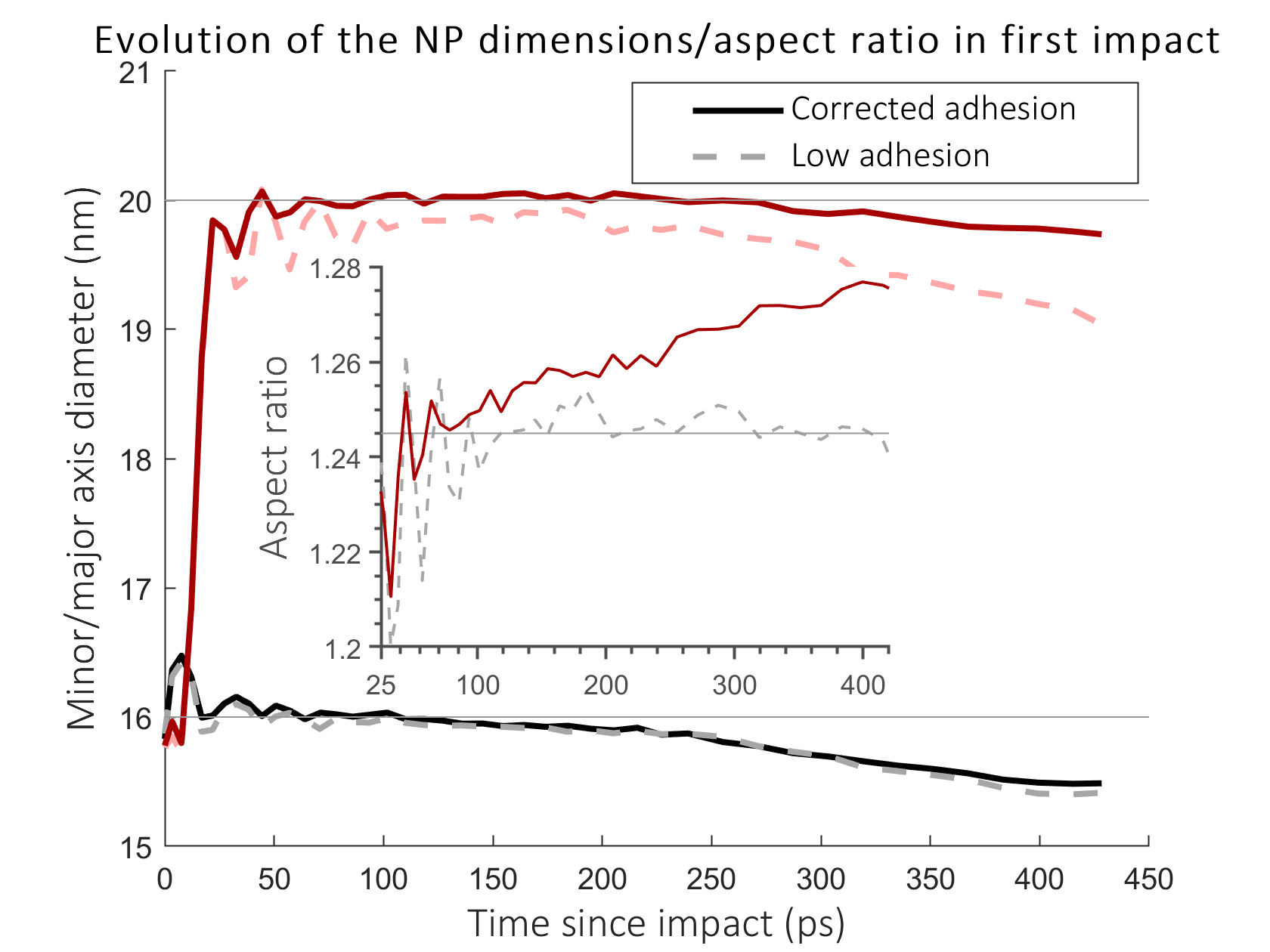}
   \label{fig:aspect_ratio_single}
}
\subfloat[]
{
    \includegraphics[width=0.5\textwidth]{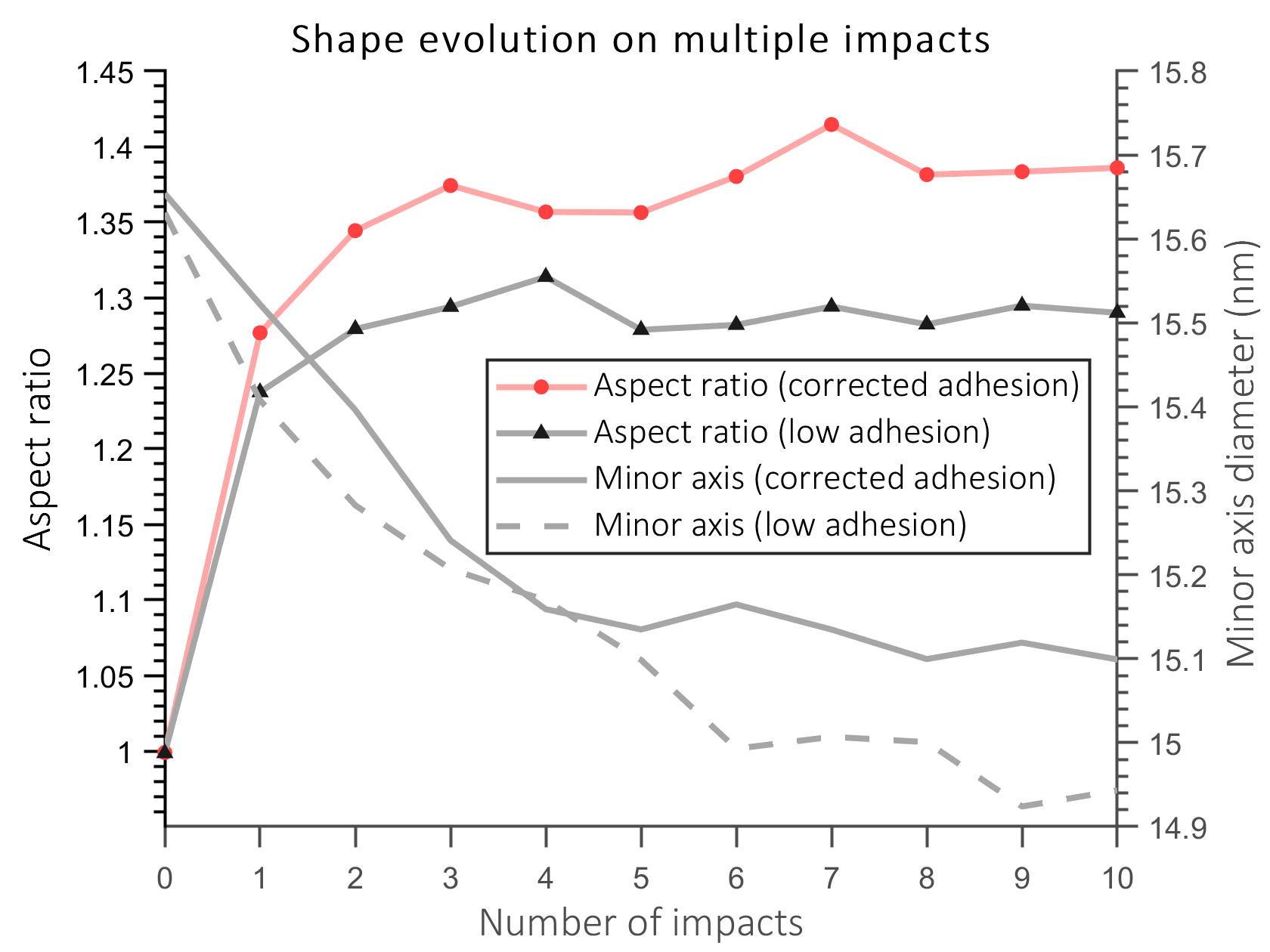}
    \label{fig:aspect_ratio_several}
}
\caption{a) NP dimension evolution in the first impact. Solid lines indicate simulations with corrected adhesive energy and dashed without. The inset shows the evolution of the aspect ratio after the ion track in silica has solidified (after about 20-30 ps). It indicates aspect ratio growth during the cooling in the simulations with corrected adhesive energy, whereas the low adhesion NP cools and shrinks isotropically without prominent changes in aspect ratio.
b) The evolution of the aspect ratio and the minor axis size on multiple impacts.
}
\end{figure}

We begin the analysis by examining the first impact in the irradiation series. In our irradiation simulations, both types of simulations indicate aspect ratio growth as shown in Fig. \ref{fig:aspect_ratio_single} and described previously\cite{Leino12, Leino14, PenaRodriguez17}. During the first 20 ps, molten, high-pressure gold pushes into the underdense ion track in silica. The initial part of the simulation, when most of the aspect ratio growth occurs, behaves identically in both simulations. However, the total aspect ratio growth is higher in the simulation with higher adhesion, and differences occur after 20 ps.  At this time, the molten ion track in silica solidifies, while the NP still remains molten. The NP and the ion track in silica cool down due to heat dissipation from the track region to bulk, but the cooling is significantly faster in silica than in the NP due to weak heat conduction through NP/silica interface. After the freezing of the ion track, no more material can flow to the ion track. Therefore, no further increment in the major axis size of NP is possible. However, the aspect ratio growth continues in the simulation with the high adhesive energy even after 20 ps.
Since the minor axis size evolves identically in both simulations \FD{(the minor axes in both cases shrink during the cooling phase), the behavior of the material along 
the major axis explains the difference in aspect ratio development during the impact and right after it,} see Fig.  \ref{fig:aspect_ratio_single}. 
The NP with the low adhesion detaches almost completely 
from the ion track in silica and shrinks \FD{during the cooling phase nearly}  isotropically. Such behavior will not lead to 
further changes in the aspect ratio 
during cooling, and small voids form instead above and beneath the particle. However, the NP with the corrected 
adhesion to the ion track resists the formation of these voids, and the major axis does not shrink in size \FD{as much as with the low adhesion.} 
As a result, the aspect ratio keeps increasing as the NP cools down. This observation demonstrates that shape change can also occur after the ion track has solidified. It is also worth noting 
that no width loss occurs during the initial part of the simulation. When the modified nanoparticle cools down, 
its elongated shape \FD{inevitably leads to further loss in its width.} 

\subsection*{Effect of multiple impacts}

\begin{figure}[ht]
\subfloat[]
{
    \includegraphics[width=0.24\textwidth]{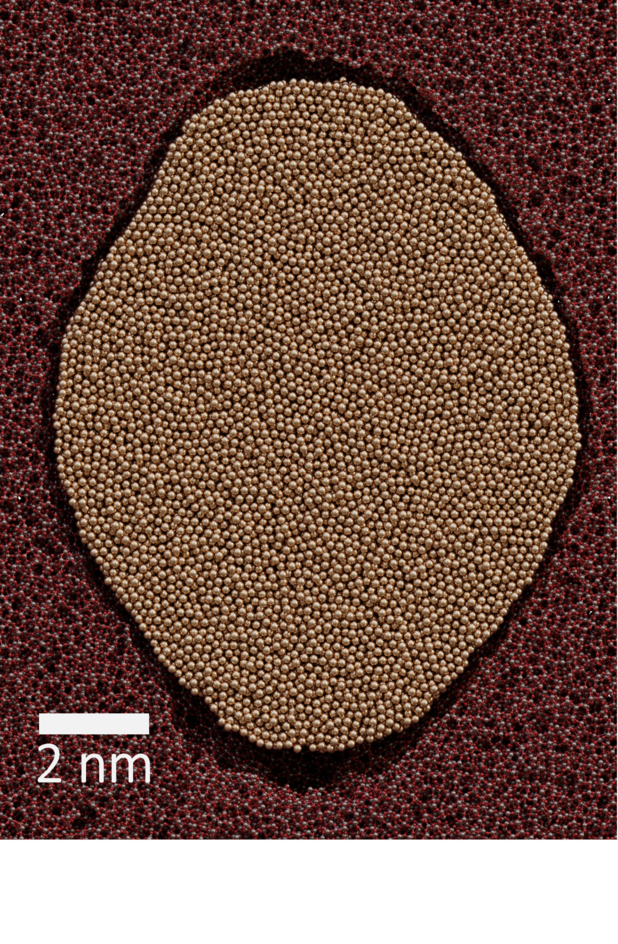}
    \label{fig:low_adhesion_10}
}
\subfloat[]
{
    \includegraphics[width=0.24\textwidth]{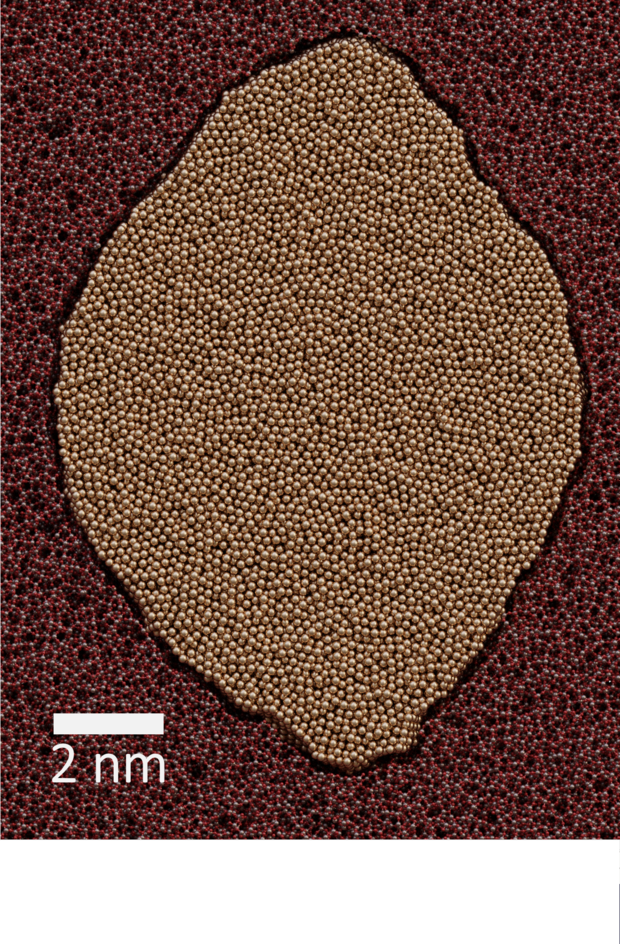}
    \label{fig:high_adhesion_10}
}
\subfloat[]
{
    \includegraphics[width=0.5\textwidth]{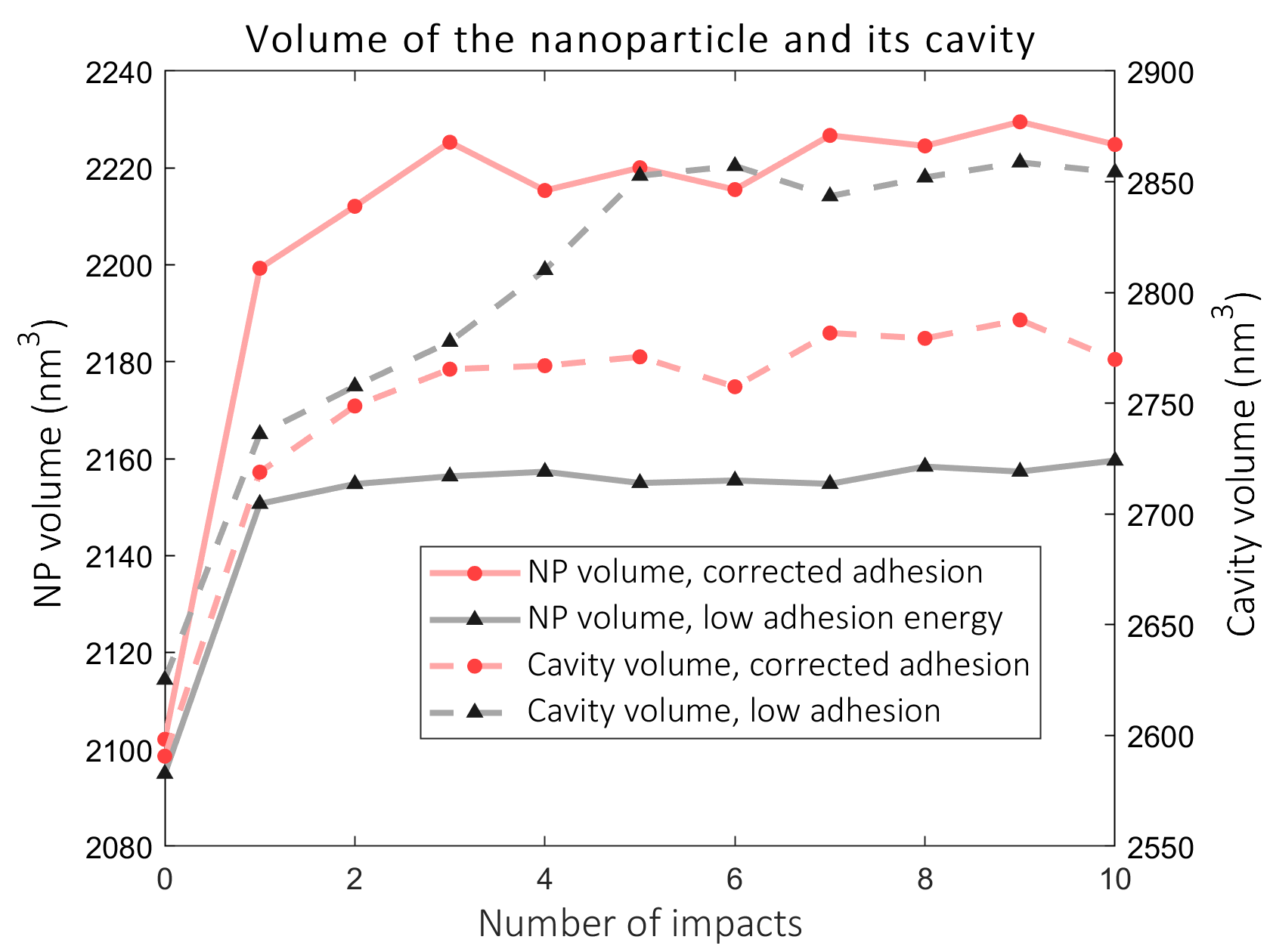}
    \label{fig:volume}
}

\caption{a) Snapshot of the cell after the 10th impact at 300 K for the simulation with low adhesion (cell cut in half). It shows the amorphous structure of the NP and an empty region around it. Larger voids can be seen on top and below the NP. b) Shown is the same for the simulation with corrected and stronger adhesion. No empty regions are visible around the NP.
c) The volume of the NP and the cavity of the NP (the void that would result by removing all Au atoms from the system) at 300K.
}
\label{fig:basic_growth}
\end{figure}

 Similar differences persist in the simulations also on later impacts. After ten impacts, the aspect ratio has grown to the values of 1.3 and 1.4 for the simulations with the low and corrected adhesion potential, respectively, as seen in Fig \ref{fig:aspect_ratio_several}. This confirms that the adhesion affects the elongation rate. Both particles continue to lose width and change shape on the first impacts but reach a state by the tenth impact in which no further, significant changes occur. The void on top and beneath the NP in the simulation with low adhesion is visible from Fig. \ref{fig:low_adhesion_10}. In addition, a 
 small gap with no atoms everywhere around the NP can be seen. The evolutions of the NP volume 
 and the NP cavity 
 \AL{(the void in silica wherein the NP resides)} 
 are shown in Fig. \ref{fig:volume}. \FD{The figure shows that the volume of both NPs 
 increases compared to the original }state 
 in the crystalline FCC configuration. \AL{We observe that the volume of the cavity and NP evolves similarly with the stronger adhesion.  With low adhesion, the cavity grows bigger, but the NP is smaller and saturates after a few impacts (note the different scale for NP and cavity volume).} 

\begin{figure}[ht!]
\subfloat[]
{
    \includegraphics[width=.24\textwidth]{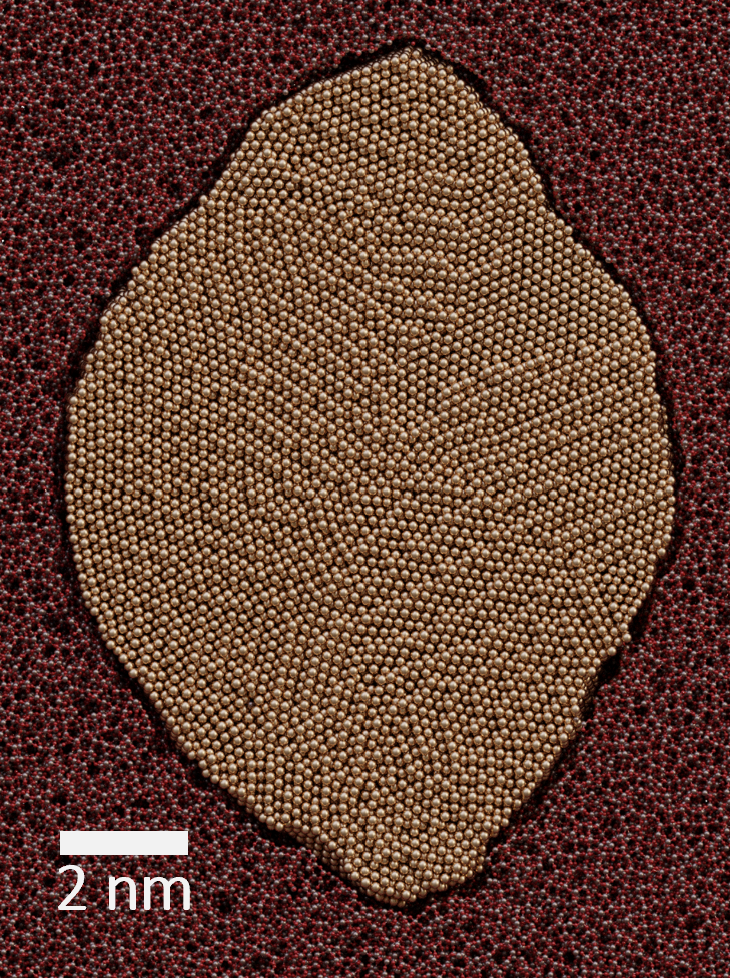}
    \label{fig:crystalline_high_adhesion_10}
}
\subfloat[]
{
    \includegraphics[width=0.37\textwidth]{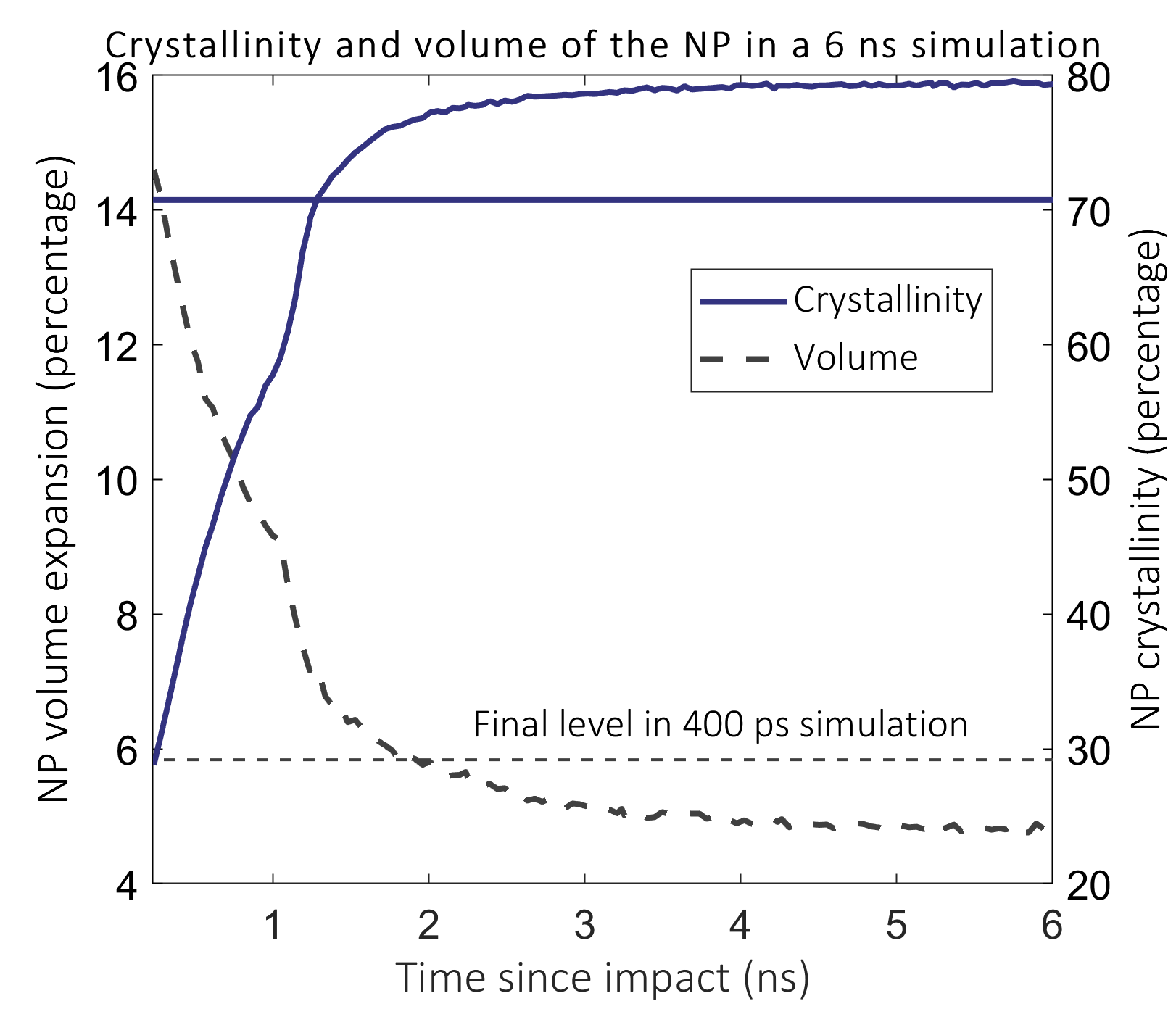}
    \label{fig:cryscon}
}
\subfloat[]
{
    \includegraphics[width=0.35\textwidth]{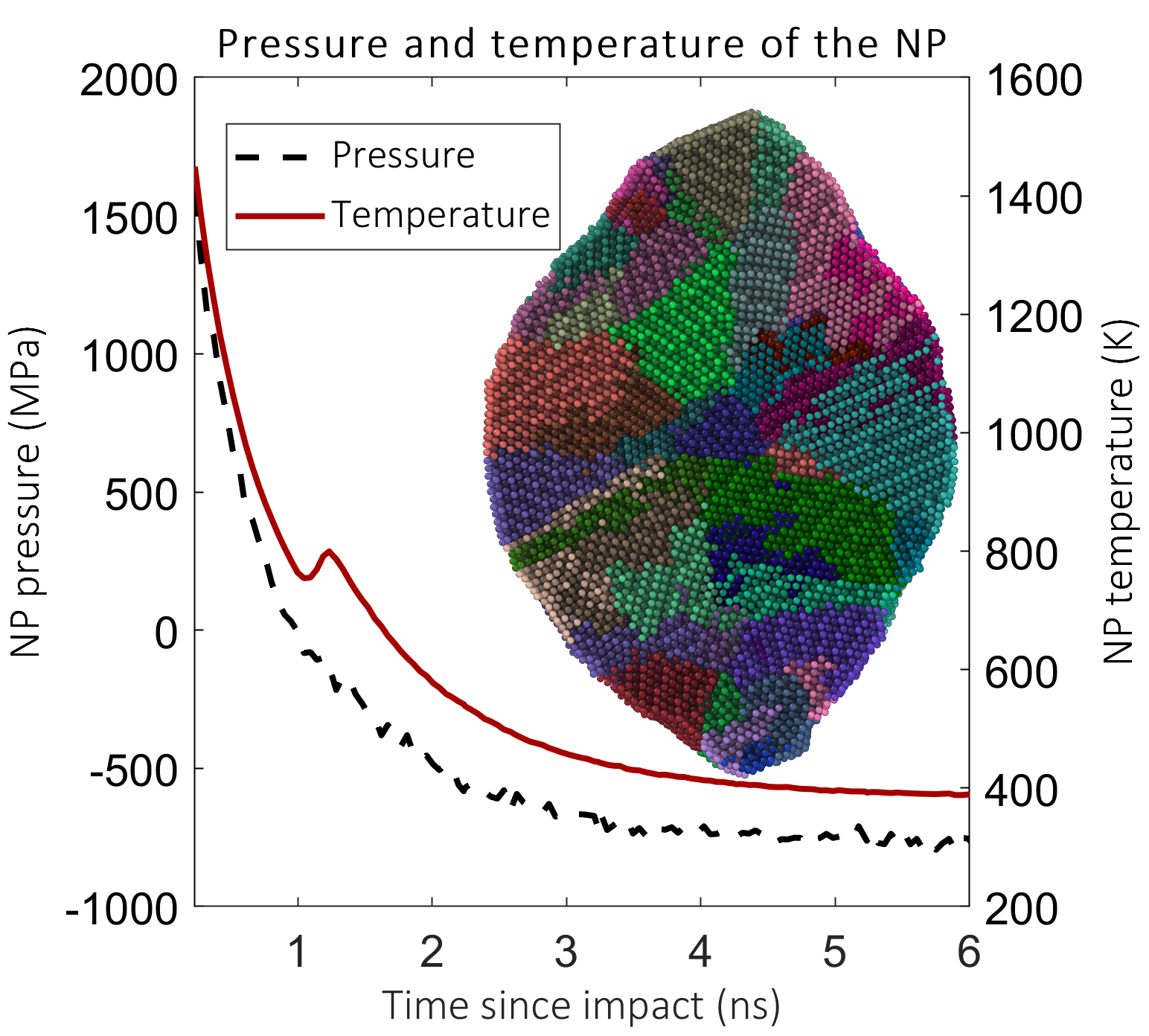}
    \label{fig:prestemp}
}

\caption{ 
a) Snapshot of the simulation cell after the 6 ns simulation. The cell is cut in half. A crystalline grain is clearly visible in the upper-right corner of the NP. Rest of the grain do in the  
b) Crystal content and the volume of the NP as a function of time in the long MD simulation. Shown by the horizontal lines are the corresponding levels in the simulations without the long MD simulation between impacts.
c) Shown is the temperature and the pressure of the NP. The inset shows colored grains that are identified using the grain segmentation tool in OVITO \cite{Stukowski10}.
}
\label{fig:longrelax}
\end{figure}

\FD{In order to reach} the experimental fluences (for example, 2 $\times$ 10${}^{14}$ ions/cm${}^2$ in Ref. \cite{Leino14}), which result in transformation of initially spherical NPs into nanorods, 
the NP must experience at least 
hundred ion impacts in our simulations. 
The impacts near the edges of the NP are not expected to raise the temperature above the melting point\cite{Dufour12}. On the other hand, rate of elongation depends on the temperature of the cluster and melting\cite{Leino12}. Since we restrict the impacts to the central part, the NP in the simulation should reach a highly elongated state with fewer impacts. However, both simulation series indicate saturation of the aspect ratio growth, as shown in Fig. \ref{fig:aspect_ratio_several}. To confirm the saturation in the case of stronger adhesion, we further extend the total number of impacts to 14. No further net change occurs (see Supplementary Material for details). The saturation was also previously observed in Refs. \cite{Leino12, Leino14}. 

We argued in Ref. \cite{Leino14} that the increased volume of the NP in later impacts decreases the volume expansion, which prevents further shape change. The saturation could be avoided only by a "forced recrystallization" - when the NP was rebuilt in the crystalline state after impact so that the shape and atom count were conserved. On the other hand, the authors in Ref. \cite{PenaRodriguez17} obtained a higher aspect ratio using a similar simulation setup without the recrystallization procedure. The main difference to the simulations presented here is the notably higher temperature (9000 K vs. 2000 K) of the NP. Due to the higher temperature, the NP can expand more on impacts, which explains why shape change was more prominent. It appeared the NP keeped on growing in volume on each impact without losing width significantly. However, due to the number conservation of metal atoms, the elongation cannot proceed only by expansion in volume. While some volume growth could occur by amorphization and more by mixing, the elongated NPs have been measured to be crystalline\cite{SilvaPereyra10}. Moreover, we observe a negligible amount of mixing of silicon and oxygen atoms to the NP in our MD simulations. Elongation by expansion without mixing should only lead to a metastable state and cause the nanoparticle to collapse due to negative pressure. To study this, we modified the simulations so that the NP obtains a temperature of 3000 K on impacts. This simulation confirmed that the NP initially elongates more but eventually collapses during cooling so that all aspect ratio growth is lost (see Supplementary Material for details).


\begin{figure}[ht!]
\subfloat[]
{
    \includegraphics[width=.5\textwidth]{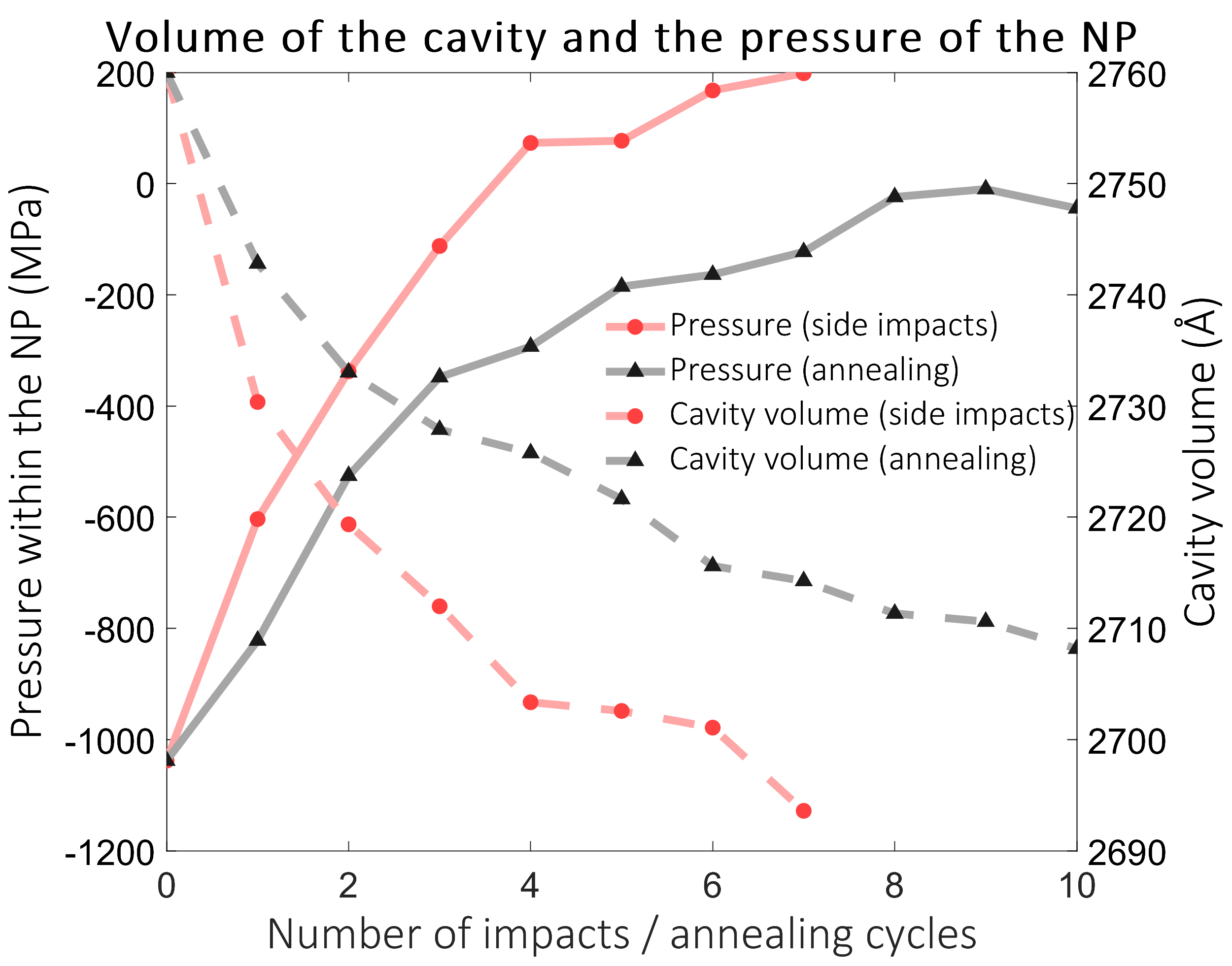}
    \label{fig:rel_comp}
}
\subfloat[]
{
    \includegraphics[width=0.45\textwidth]{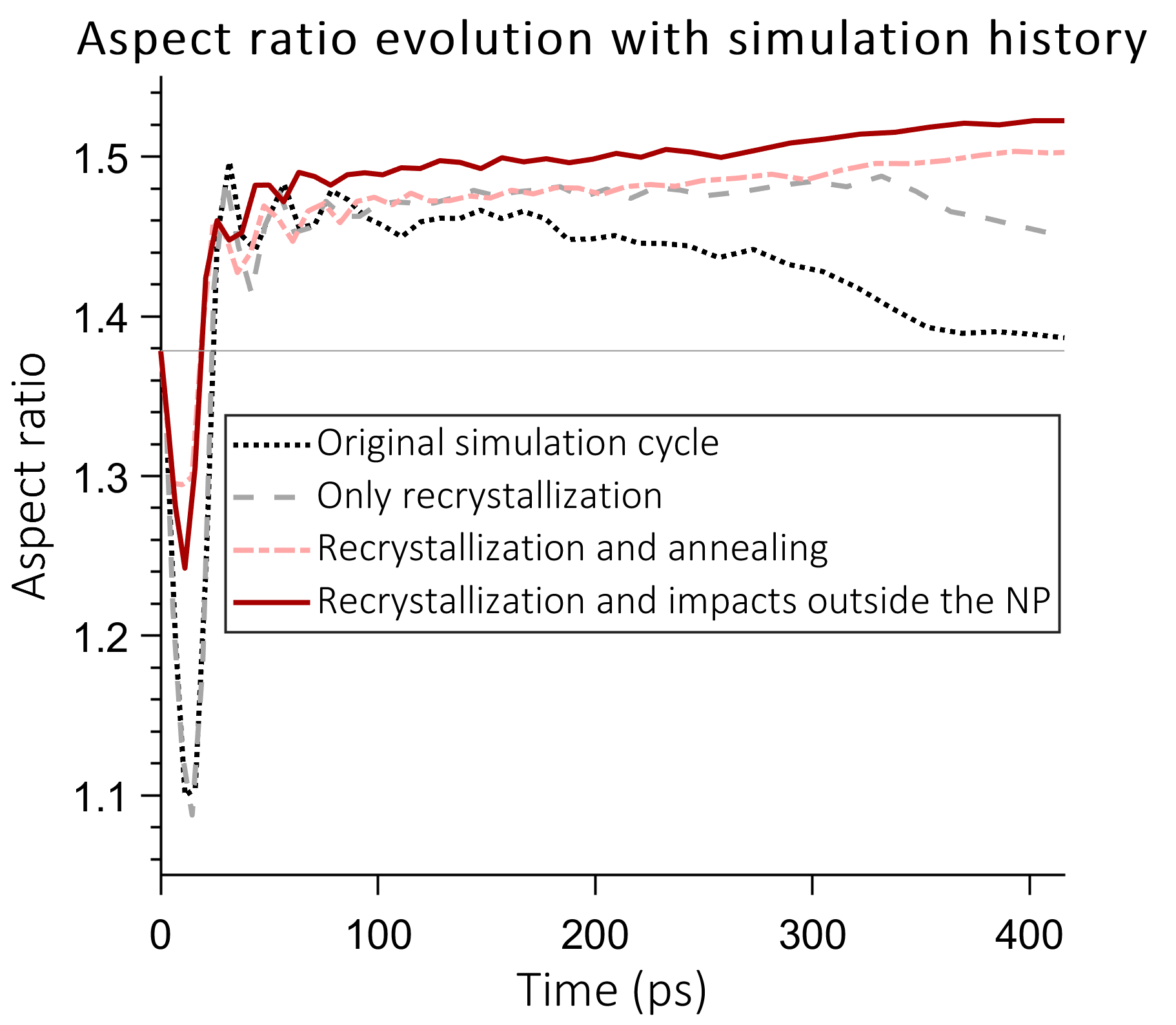}
    \label{fig:ar_comp}
}

\caption{ 
a) The pressure and the cavity volume (void that would result by removing Au atoms) during impacts nearby. Also shown are the same results from an alternative annealing treatment.
b) Comparison of aspect ratio growth on the 11th impact using simulation cells with different histories.
}
\label{fig:relaxelongation}
\end{figure}
\subsection*{Longer time scale evolution}

To shed light on the importance of recrystallization without enforcing it artificially, we study a scenario in which the NP is let to relax for a longer time between impacts 
in MD simulations. We expect that the recrystallization provides a mechanism by which the NPs lose volume, and elongate during the next impact\cite{Leino14}. However, we note that with low adhesion, this mechanism might not reduce the volume of the cavity as silica and Au are not strongly coupled. Empty region around the NP is visible from \ref{fig:low_adhesion_10} and can be verified from Fig. \ref{fig:volume}. Therefore, even if the NP reverts to the original volume, it might only expand to the enlargened cavity during the next impact without a further net change in aspect ratio. The forced recrystallization process introduced in Ref.\cite{Leino14} bypasses this issue since the size of the cavity is automatically adjusted to the size of the NP. In the case of high adhesion, the cavity is smaller (see Fig. \ref{fig:volume}), but the NP itself is in a higher volume state. Our goal is to study if long-timescale simulations will bring the particle and the cavity size down due to strong coupling so that the elongation can proceed as explained previously. Due to the high computational \FD{costs} 
of the simulation and improved realism, we choose to simulate the long-term processes only on the NP with corrected adhesion.

During the previous simulation cycle, the NP evolves for 230 ps after the impact with boundary conditions that mimic heat conduction to the bulk. Another 200 ps is reserved to cooling the NP to 300 K and pressure relaxation. Instead of the previous 200 ps run, we let the NP cool down with only boundary cooling for six ns on the 10th impact. The crystallinity, volume, temperature and pressure of the NP during the simulation are shown in Fig. \ref{fig:longrelax}. 
Fig \ref{fig:crystalline_high_adhesion_10} shows a snapshot of the simulation at the end. It is evident that crystalline grains form and, as argued previously, the shape of the particle does not change.
Fig. \ref{fig:cryscon} shows that the particle reaches about 80\% crystallinity, but the NP is still about 4\% over its original volume at the end of the simulation. Fig. \ref{fig:prestemp}
 shows the evolution of the temperature and pressure during the relaxation. Initially, a large overpressure ($>$ 1.5 GPa) forms in the NP in line with previous arguments. However, as the particle cools down, a significant underpressure (-0.8 GPa) develops in the cluster. Even after cooling down, the volume of the NP is higher than that of the NP with low adhesion in the amorphous state. Moreover, the final cavity size does not change compared to the shorter simulation. Hence we conclude that another mechanism is required to bring the size of the cavity down. Furthermore, the adhesion between the NP and the cavity walls causes considerable underpressure in the NP, which will decrease the initial pressure on subsequent impacts.


A missing element in the simulations is ions hitting near the NP but not intersecting it directly. Bombarding the simulation cell at random positions with no respect to the NP location improves the realism of the simulation, however, only about 16\% of the ions would hit the NP. To study the effects of impacts outside the NP, we continue the previous 6 ns simulation with bombardment near the NP. The ions hit in a circular region around the NP at a radius that is 2 nm larger than the NP, as illustrated in Fig. \ref{fig:kirkas}. The volume of the NP and the pressure within it after each impact are shown in Fig \ref{fig:rel_comp}. After 7 impacts, the negative pressure within the NP has decreased and the width of the NP has reduced. \AL{The simulation cell shows characteristics of the compactification preceding the ion hammering effect: the density of silica has increased by about 10\%.} The simulations show that the cavity size decreases by the nearby impacts. 

However, irradiating the cell in this manner is still computationally prohibitively expensive. As an alternative method to mimic the evolution of cell under thermal spikes, we anneal the simulation cells and control the pressure. In the annealing cycle, the temperature of the cell is increased linearly to a target temperature and cooled back to room temperature. The heating rate (and cooling, 120 K/ps) and the target temperature and pressure (3500 K, 0 GPa) are chosen so that there is not enough time for melting to occur, but atoms gain enough kinetic energy to find new equilibrium positions. As seen from Fig. \ref{fig:rel_comp}, about 8 annealing cycles remove the negative pressure from the NP. We note during this process, the ion track structure (density changes induced by the ion) in silica is preserved, and the median displacement of atoms during a single annealing cycle is only 0.4 Å. The median after nine annealing cycles is 3.6 Å.
\begin{figure}[ht!]
\subfloat[]
{
    \includegraphics[width=.33\textwidth]{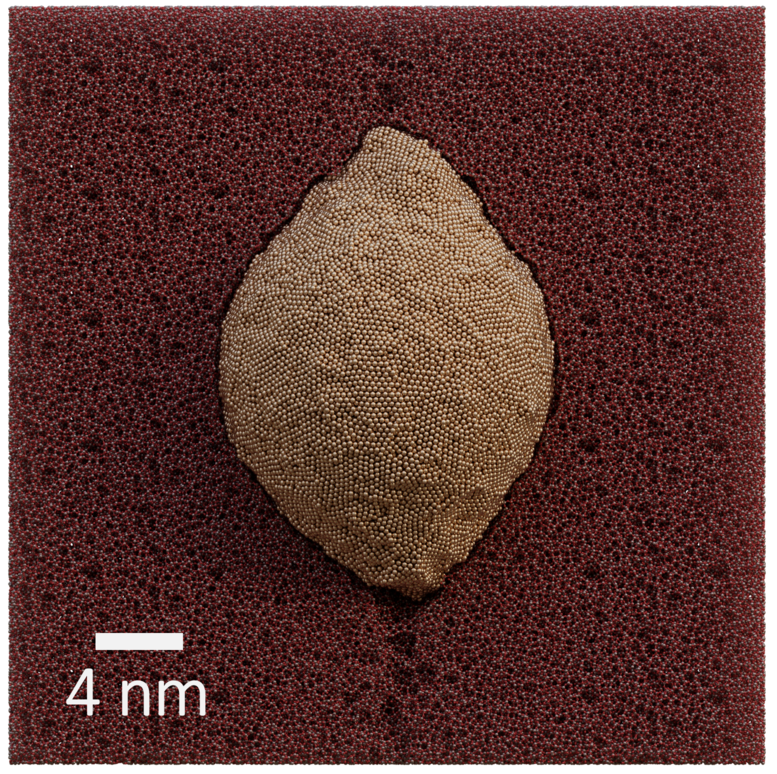}
    \label{fig:relaxelongation_a}
}
\subfloat[]
{
    \includegraphics[width=0.33\textwidth]{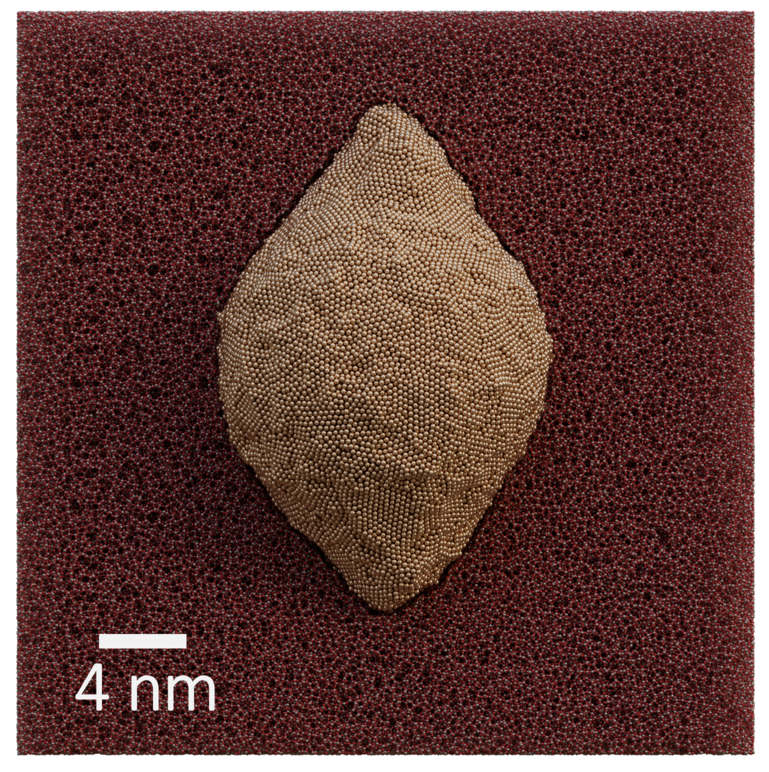}
    \label{fig:relaxelongation_b}
}
\subfloat[]
{
    \includegraphics[width=0.33\textwidth]{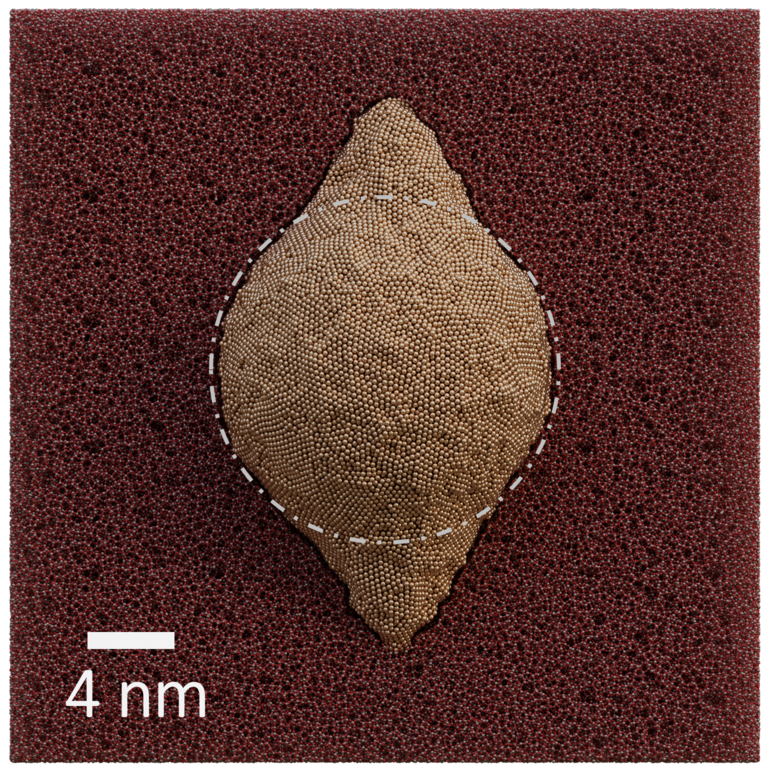}
    \label{fig:relaxelongation_c}
}

\caption{
a) Shown is the NP after the 11th impact using the original simulation cycle (the NP is not cut in half)
b) As previous, but with 6 ns simulation and impacts outside the NP
c) NP after 12th impact using the alternative the annealing method on impacts 10 and 11. Indicated by the white dashed line is the shape of the NP at the start of the simulation. 
}
\label{fig:finalimages}
\end{figure}
\subsection*{Effect of the longer timescale evolution on the subsequent impacts}
\AL{
Finally, we consider how the relaxation processes affect the evolution of the aspect ratio on subsequent impacts. A simulation is performed in cells with different histories so that the ion hits the central axis of the NP. Figs. \ref{fig:ar_comp} and \ref{fig:finalimages} show the results from the comparison. Simulation with the original cycle shows negligible aspect ratio growth (see Figs. \ref{fig:ar_comp} and \ref{fig:relaxelongation_a}). The six ns recrystallization simulation increases the aspect ratio growth compared to the original simulation cycle only modestly. A clear difference occurs when the long recrystallization simulation is continued with nearby impacts (Figs. \ref{fig:ar_comp}, \ref{fig:relaxelongation_b}) or when using the thermal annealing approach instead. We have also perform a second subsequent impact with the annealing process. In this test, the NP is relaxed again with a six ns simulation for recrystallization and the cell again annealed. A prominent aspect ratio growth occurs again from the initial value of 1.5 to a final aspect ratio of 1.64 (note that the corresponding aspect ratio is 1.38 in the original cycle). The NP is shown in Fig. \ref{fig:relaxelongation_c}. Therefore, we conclude that the saturation is an artificial effect not manifesting in a more realistic simulation setting. Moreover, the figure shows that the NP has lost width, although the decrease is only $\sim$1 nm due to the small number of impacts. A similar diagram for the snapshots presented in earlier works\cite{MotaSantiago17, Leino12} indicates expansion only.}


\section*{Conclusions}

In summary, we have shown that the elongation process can be reproduced in MD simulations only when the effects of ions that do not hit the NP are taken into account. It is also necessary to include an accurate description of adhesive forces between gold and silica as they directly affect the elongation rate. Both observations are critical for creating theoretical predictions of NP shape evolution with fluence. Moreover, the results have significance for understanding the fundamentals of ion-solid interactions, as they show that the elongation can occur at lower NP temperatures than used earlier work\cite{MotaSantiago17}, similar to those predicted using the two-temperature model\cite{Dufour12}.

 These results may be used as a starting point to understand seemingly incompatible experimental observations: elongation induced by individual impacts to NPs\cite{amekura2017, Rizza12} but the overall process influenced by the embedding matrix\cite{Roorda04, Coulon2016, Li2020}. The results show unambiguously that nearby impacts affect the stress within the NP, which changes the elongation rate. The previous MD simulations\cite{Leino12,Leino14, MotaSantiago17} indicated a more passive role of the embedding matrix. However, elongation is induced already by the first impact on the NP, demonstrating that the effect can occur without a threshold fluence, which is consistent with experimental observations\cite{amekura2017}. \AL{The primary mechanism for elongation is consistent with the molten material flow to the underdense ion track in silica. \cite{Leino14}. This flow is due to the overpressure in the metal NP as suggested previously\cite{DOrleans03}. However, the NP recrystallizes in a highly stressed state which must be relaxed before further, significant shape modification can occur. }

\AL{
Recently, the first experimental investigations studying the dependence of the effect on the embedding matrix were reported\cite{Li2020, MotaSantiago17}. We believe that our results will be important to interpret the experiments, as they show that the metal-matrix interactions affect the elongation rate. We have demonstrated the process to be more complex than thought earlier, and a lot remains to be explored. Despite already a long history of experimental research, only a handful of papers address the shape transformation problem theoretically. We hope that our work inspires and paves the way for further analysis. The insights presented here are a step toward accurate simulations and controlled shaping of the NPs experimentally. 
}

\section*{Methods}

\subsection*{Interatomic potentials}
 \label{sec:potential}
  Si-O, Si-Si, and O-O interactions were described by the Munetoh potential\cite{Munetoh2007}, which is of Tersoff-type. Au-Au interactions were depicted by an EAM-potential by Foiles et al.\cite{Foiles86}. For Au-Si, Au-O interactions, we developed a pair potential fitted to first principles calculation made with the Gaussian 16\cite{g16} software.

We used the DFT method with the  Becke 3-parameter Lee-Yang-Parr (B3LYP) hybrid functional and a mixed, localized basis set. On Si and O atoms, Pople-style basis set with augmented with diffuse functions (6-311++G(d,p)) was used, whereas for the gold atom, Los Alamos National Laboratory 2-double-z (LANL2DZ) basis set was used. Similar setup was used, for example, by Del Vitto et al.\cite{DelVitto05} and Lu et al.\cite{Lu18} 

\subsubsection*{Computation details}

We first relaxed the structure so that the gold atom lies outside of the molecule, at the symmetry axis (x-axis). Potential energy surface scan was then performed along the symmetry axis as shown in Figure \ref{fig:pes_scan}. The total energy at x=15 nm was chosen as the zero level energy, and subtracted from the DFT results. Origin was chosen so that x=0 nm means gold is in the middle of the molecule. To express the total energy from the DFT computation as individual pair contributions from Morse potentials, the total energy was evaluated as
\begin{equation}
\begin{split}
E_{\text{Morse}}(x) = \sum_{i=\text{Si}_1...\text{Si}_8 - \text{Au}} D_{e,\text{Si-Au}} ( e^{-2a_{\text{Si-Au}}(r_i(x) - r_{e,\text{Si-Au}})} - 2e^{-a_{\text{Si-Au}}(r_i(x) - r_{e, \text{Si-Au}})} ) \\
+\sum_{i=\text{O}_1...\text{O}_{12} - \text{Au}} D_{e,\text{Si-O}} ( e^{-2a_{\text{Si-O}}(r_i(x) - r_{e,\text{Si-O}})} - 2e^{-a_{\text{Si-O}}(r_i(x) - r_{e, \text{Si-O}})} )
\end{split}
\label{eqn:fit}
\end{equation}
where the summations go over all gold atom pairs in the system, and $r_i(x)$ is the radius of an individual bond when the gold atom lies at $x$. $D_e$, $a$ and $r_e$ pairs for Au-Si and Au-O are the well depths, bond widths and equilibrium distances for each type of interaction. These six parameters were optimized by minimizing the squared difference between the energies obtained from DFT potential surface energy scan and those evaluated using Equation \ref{eqn:fit} at the scan locations (shown by the dots in Fig. \ref{fig:pes_scan}).
 The Jacobian matrix with respect to the parameters was formed using analytic differentiation, and the mimization was performed using a Gauss-Newton algorithm. The resulting parameters are given in Table 1 and a comparison between the fitted and DFT potential energy in Fig. \ref{fig:pes_scan}. 

We also computed corrections to the potential based on the experimental adhesive energy. To compute the adhesion energy predicted by the fitted potential, we prepared a spherical Au NP in cubic a silica simulation cell as described in section \textit{Simulation procedures}. The NP radius was 8 nm. Basing the computations on the the relaxed room temperature system as the  (300 K, 0 GPA), we estimated the adhesive energy as (following Ref.\cite{Huhn15}) 
\begin{equation}
E_{\text{adh}} = \frac{E_{\text{Au}} + E_{\text{SiO}{}_2} - E_{\text{Au/SiO}{}_2}}{4 \pi r^2}
\end{equation}
where $r$ is the NP radius, $E_{\text{Au}}$ and $E_{\text{SiO}{}_2}$ refer to the total energies of the gold nanoparticle and the cubic simulation box separately, and $\text{Au/SiO}_2$ to the energy of the composite system.

To fit the adhesive energy, we multiplied the potential energy values given by the original fit on the right hand side of the potential well minimum with different scaling factors (2,4,6,10). The potential energy values far from the minimum were kept unchanged (values in the range $x \lessapprox 4 $ Å in figure \ref{fig:pes_scan}), and the remaining ones removed. A new fit to the resulting potential energy values for all the scaling factors was performed. The resulting new potentials were used to construct the NP in silica system from the beginning. It was observed that using value of four as the scaling factor gave the closest adhesive energy value to the experimental one.

\subsection*{Simulation procedures}
\label{sec:simulation}

All MD simulations are performed using the LAMMPS\cite{Plimpton95} software package (http://lammps.sandia.gov). 

\subsubsection*{Initial cell for impact simulations and adhesive energy computations}
A smaller cubic simulation cell with a side length of 5.8 nm was obtained with the WWW-method as described elsewhere\cite{Alfthan03}. It was then multiplied six times for a larger silica cell (35 nm x 35 nm x 35 nm) and relaxed with the Munetoh potential. A sphere with a diameter d=16 nm was cut from FCC Au,  compressed by 1\%, and inserted into a spherical cavity in the silica cell, which was cut slightly larger (d=16.2 nm). Next, the system was equilibrated for 100 ps towards 0 GPa/300 K using a Langevin thermostat and a 0.8 fs time step (thermostat time constants were 80 fs for temperature and 800 fs for pressure). The purpose of this procedure is to make the initial cell stress-free. Moreover, the compression forces the nanoparticle to interact with silica without low interatomic separations in the initial configuration that might result in atoms with high velocity near the surface.

\subsubsection*{Impact simulation and the main simulation cycle}

Schematics of the simulation cell configuration for impact simulation is given in Fig. \ref{fig:schematics}. To model the rapid heating of atoms following a swift heavy ion impact, random velocity vectors are added instantaneously so that the kinetic energy increase corresponds to that estimated from the two-temperature model for SiO${}_{\text{2}}$. For gold, we have tried both instantaneous and heating the NP to 2000 K in 5 ps duration to account for the slower electron-phonon interactions.
We chose use the latter approach for the main simulation cycle, as it increases the realism according to the two-temperature model. The value of 2000 K is chosen on qualitative grounds, but in line with earlier estimates using the two-temperature model with the Au NP-silica system\cite{Awazu09, Dufour12} and similar ions. It is enough to melt gold (melting point with the EAM potential is 1110 $\pm$ 20 K \cite{Leino14}) but enough not to vaporize it. When the ion does not hit the nanoparticle (see Fig. \ref{fig:kirkas}), a temperature of 600 K is used instead, which is similarly in line with earlier estimates\cite{Dufour12}.
The instantaneous radial deposition profile in SiO${}_{\text{2}}$ is based on 164 MeV Au ion and the as same used in Refs. \cite{Leino12,Leino14} and is given in the Supplementary Material for completeness. 

The simulation cell is cooled at four boundaries parallel to the ion direction to dampen pressure waves and to mimic heat conduction to bulk. The width of the boundary cooling region is 1 nm and the cooling is performed using a Berendsen thermostat with a rapid time constant (10 fs) towards 300 K. Once the heating of Au is over, simulation continues with the boundary thermostat still turned on for 230 ps. This stage is followed by a pressure/temperature relaxation with a total duration of 200 ps using a time constant of 100 ps in the Berendsen barostat and thermostat. The thermostat is applied separately to all SiO${}_2$ and Au atoms to ensure that both systems are at 300 K before the next impact (and not only the composite system). After this relaxation, shifting over periodic boundaries is performed as explained in the main text, and the simulation cycle is started from the beginning for more impacts to the cluster.


\subsubsection*{Nanoparticle characterisation}
The dimension and the aspect ratio were computed using an average position of 100 furthest atoms in each dimension. The volume of the nanoparticle was determined using the coordinates of gold atoms and the convex hull volume determination in MATLAB (The MathWorks Inc., Natick, MA). For quantifying the size of the cavity, Voronoi analysis was used to determine the first silica neighbors of gold atoms. We determined the convex hull volume of these first neighbors, and computed its volume. To characterize the crystal content in the nanoparticle, we used the Ackland-Jones analysis\cite{Ackland06} as implemented in the OVITO software\cite{Stukowski10}. The percentage of crystalline content was computed as the ratio of atoms in any crystal structure (FCC, HCP, BCC, ICO) over unidentified atoms.

\bibliography{main}

\section*{Competing interests}
The authors declare no competing interests.

\section*{Acknowledgements}
We acknowledge CSC – IT Center for Science, Finland for providing computational
resources. The simulations also utilized the Finnish research infrastructure urn:nbn:fi:research-infras-2016072533. VJ acknowledges funding by the NANOIS project of the Academy of Finland.


\end{document}


\flushbottom
\maketitle

\thispagestyle{empty}

\section{Saturation of growth using more impacts}

\begin{figure}[ht!]
\centering
\includegraphics[width=0.8\linewidth]{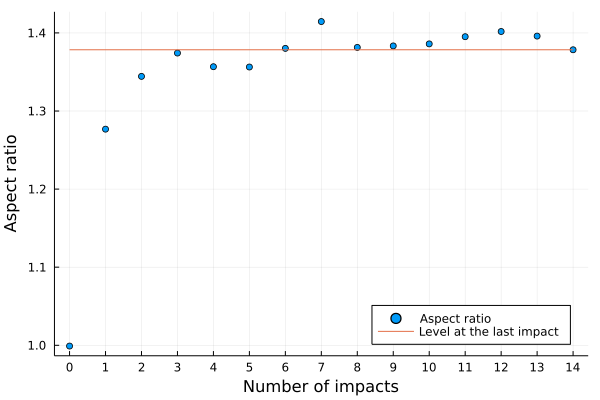}
\caption{The evolution of the aspect ratio for 14 impacts with the corrected adhesion.}
\label{fig:saturation}

\end{figure}

Figure 3b in the manuscript shows the evolution of the aspect ratio up to ten impacts. It might appear from the graph that the NP with corrected adhesion is still slowly increasing in aspect ratio. To show that this is not the case, the evolution during four more impacts is presented in Fig. \ref{fig:saturation}.

\begin{figure}[ht!]
\centering

\subfloat[]
{
\includegraphics[width=0.3\linewidth]{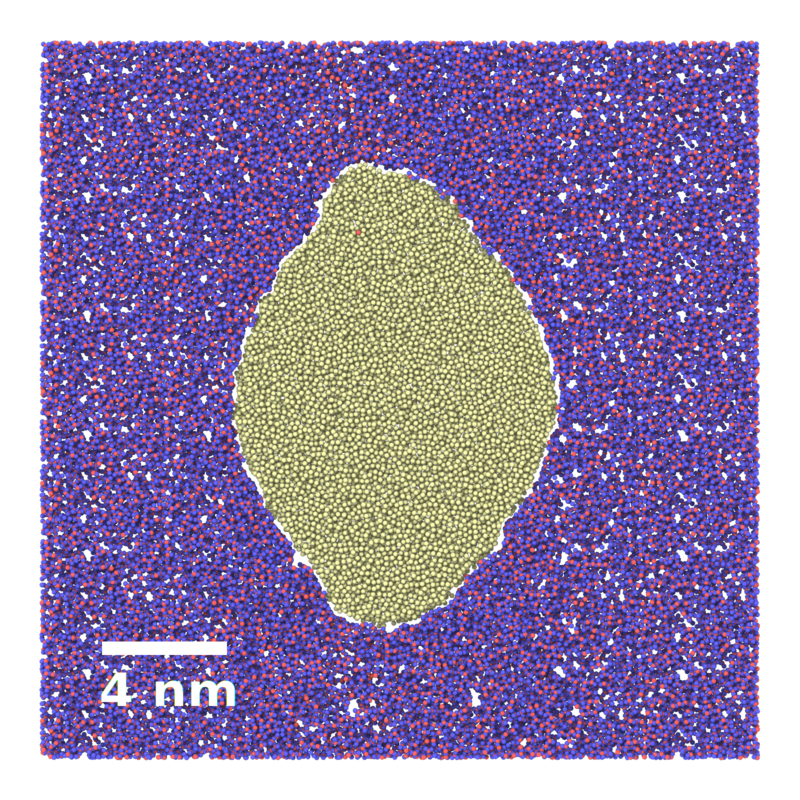}
   \label{fig:fig2}
}
\subfloat[]
{
\includegraphics[width=0.3\linewidth]{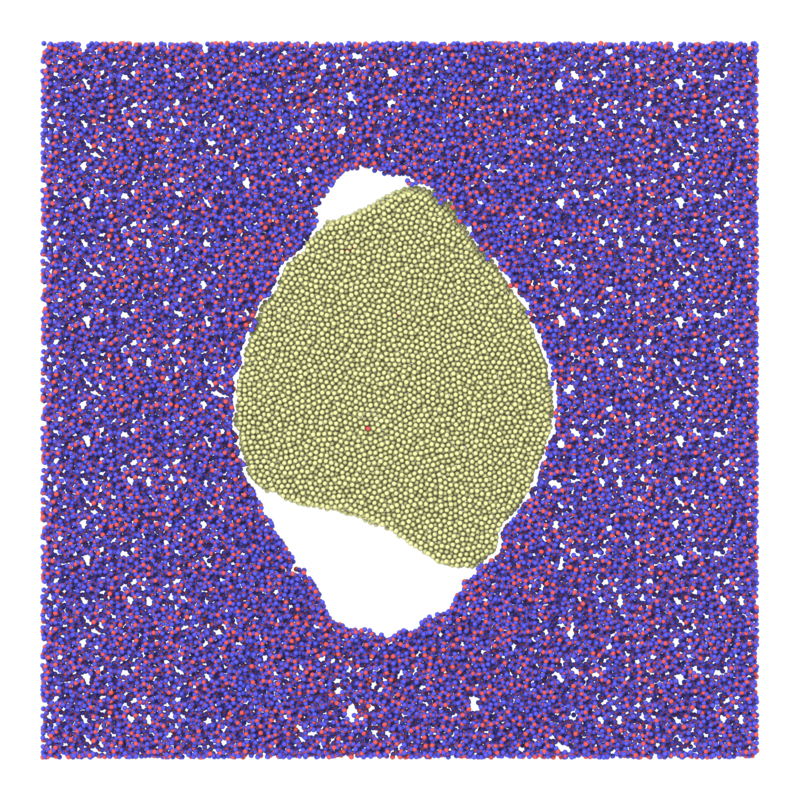}
    \label{fig:kirkas}
}
\subfloat[]
{
\includegraphics[width=0.3\linewidth]{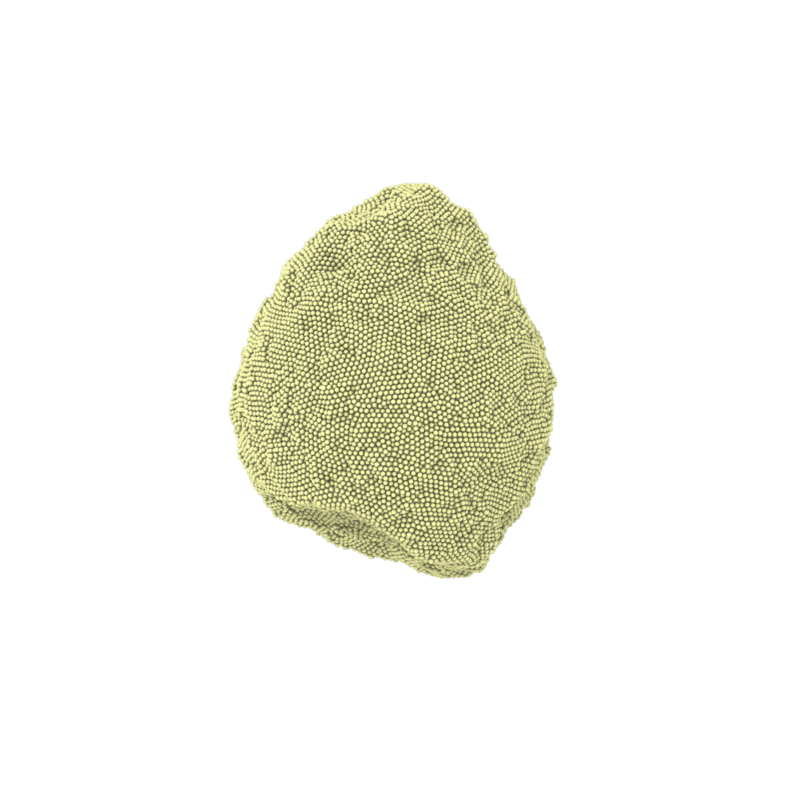}
    \label{fig:kirkas}
}

\label{edep}
\caption{Collapse of the nanoparticle at 3000 K. a) Cross sectional snapshot of the simulation cell before the collapse b) after the collapse c) All atoms of the NP after the collapse. }
\end{figure}
\section{Collapse of the nanoparticles at elevated temperatures}

In the section "Effect of multiple impacts," it is mentioned that the nanoparticles collapse when a heating temperature of 3000 K is used instead of 2000 K. Also, in this test, a heating time of 2.5 ps was used over the 5 ps time that was use elsewhere.  The collapse occurred during the 8th impact on the NP. Snapshots before and after are shown in Fig. 2. The collapse occurs during the cooling of the cell and can be explained by the high volume expansion NPs. Since the expanded volume of the NP is higher, the negative pressure accumulating in the cluster, which causes the collapse. 

\section{Energy deposition profile in silica}

The energy deposition to silica was the same as used in Refs. \cite{Leino2012, Kluth2008, Pakarinen2009} and shown in Fig. \ref{edep}. This profile is based on the so called inelastic thermal spike model and was provided to us by M. Toulemonde.
It was computed for 164 MeV Au ion in silica and had previously given a track diameter in agreement with experiments 
\cite{Kluth2008, Pakarinen2009}.  See, for example, Ref. \cite{Rotaru2012} for the parameters and the equations. The profile results from a coupled set of heat diffusion equations at a time instance when the lattice temperature has reached maximum. This occurs shortly ($\sim$ 100 fs) after the impact. As discussed in Refs. \cite{Leino2014, Kluth2008}, the the resulting profiles from the inelastic thermal spike model has been multiplied by a factor of two. This discrepancy was attributed to the inability of the potential (there, Watanabe-Samela potential  \cite{Leino2014}) to describe the melting point accurately. We note that for the purposes of the current study, this profile is only used in qualitative manner. However, using the same profile (i.e. multiplied by 2) with the Munetoh potential (used in the current work) gives quantitatively similar features for the ion track (i.e. the ion track with similar radius and underdense core). Since these are dependent on the melting point, the similarity indicates that the melting point is  overestimated also by the Munetoh potential.

\begin{figure}[ht!]
\centering
\includegraphics[width=0.8\linewidth]{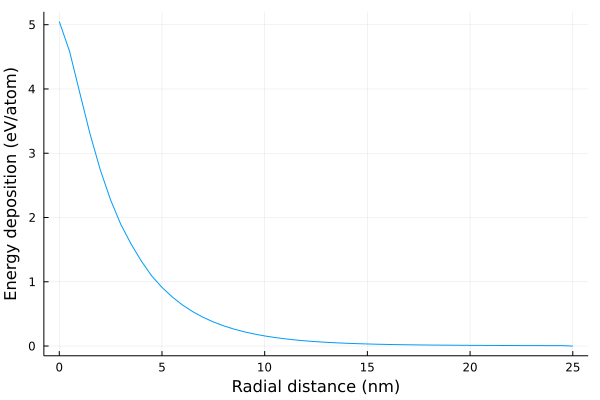}
\caption{Energy deposition to silica as a function of radius.}
\label{edep}
\end{figure}

\bibliography{supplementary}